\documentclass[letter,longauth]{aa} 
\usepackage{graphicx}
\usepackage{txfonts}
\usepackage[figuresright]{rotating}
\usepackage{natbib}
\usepackage{url}
\usepackage{color}
\usepackage{hyperref}
\usepackage{xspace}
\usepackage{soul}
\usepackage{appendix}

\makeatletter
\renewcommand*\aa@pageof{, page \thepage{} of \pageref*{LastPage}}

\newcommand{\xspec}{{\sc xspec}\xspace}

\newcommand{\MAXI}{{MAXI}\xspace}
\newcommand{\IXPE}{{\em IXPE}\xspace}
\newcommand{\Swift}{{\em Swift}\xspace}

\newcommand{\XTE}{{\em RXTE}\xspace}
\newcommand{\SAX}{{\em BeppoSAX}\xspace}

\newcommand{\xtej}{{XTE~J1701$-$462}\xspace}

\begin{document}

\title{Discovery of strongly variable X-ray polarization in the neutron star low-mass X-ray binary transient \xtej}

\titlerunning{X-ray polarization of \xtej}

\authorrunning{M. Cocchi et al.}


\author{
Massimo Cocchi \inst{\ref{in:INAF-OAC}}
\and Andrea Gnarini \inst{\ref{in:UniRoma3}}
\and Sergio Fabiani \inst{\ref{in:INAF-IAPS}}
\and Francesco Ursini \inst{\ref{in:UniRoma3}}
\and Juri Poutanen \inst{\ref{in:UTU}} 
\and Fiamma Capitanio \inst{\ref{in:INAF-IAPS}} 
\and Anna~Bobrikova \inst{\ref{in:UTU}} 
\and RubenFarinelli\inst{\ref{in:INAF-Bol}} 
\and Adamantia Paizis\inst{\ref{in:INAF-MI}} 
\and Lara Sidoli\inst{\ref{in:INAF-MI}}
\and Alexandra Veledina \inst{\ref{in:UTU}} 
\and Stefano Bianchi \inst{\ref{in:UniRoma3}}  
\and Alessandro Di Marco \inst{\ref{in:INAF-IAPS}} 
\and Adam Ingram \inst{\ref{in:Newcastle}} 
\and Jari J.~E.~Kajava \inst{\ref{in:UTU},\ref{in:ESAC}}
\and Fabio La Monaca \inst{\ref{in:INAF-IAPS}} 
\and Giorgio Matt  \inst{\ref{in:UniRoma3}}  
\and Christian Malacaria \inst{\ref{in:ISSI}}
\and Romana Miku{\v s}incov\'a \inst{\ref{in:UniRoma3}}
\and John Rankin \inst{\ref{in:INAF-IAPS}} 
\and Silvia Zane  \inst{\ref{in:MSSL}}
\and Iv\'an Agudo \inst{\ref{in:CSIC-IAA}}
\and Lucio A. Antonelli \inst{\ref{in:INAF-OAR},\ref{in:ASI-SSDC}} 
\and Matteo Bachetti \inst{\ref{in:INAF-OAC}} 
\and Luca~Baldini  \inst{\ref{in:INFN-PI},      \ref{in:UniPI}} 
\and Wayne H. Baumgartner  \inst{\ref{in:NASA-MSFC}} 
\and Ronaldo Bellazzini  \inst{\ref{in:INFN-PI}} 
\and Stephen D. Bongiorno \inst{\ref{in:NASA-MSFC}} 
\and Raffaella Bonino  \inst{\ref{in:INFN-TO},\ref{in:UniTO}}
\and Alessandro Brez  \inst{\ref{in:INFN-PI}} 
\and Niccol\`{o} Bucciantini \inst{\ref{in:INAF-Arcetri},\ref{in:UniFI},\ref{in:INFN-FI}} 
\and Simone Castellano \inst{\ref{in:INFN-PI}}  
\and Elisabetta Cavazzuti \inst{\ref{in:ASI}} 
\and Chien-Ting Chen \inst{\ref{in:USRA-MSFC}}
\and Stefano Ciprini \inst{\ref{in:INFN-Roma2},\ref{in:ASI-SSDC}}
\and Enrico     Costa \inst{\ref{in:INAF-IAPS}} 
\and Alessandra De Rosa \inst{\ref{in:INAF-IAPS}} 
\and Ettore     Del Monte \inst{\ref{in:INAF-IAPS}} 
\and Laura      Di Gesu \inst{\ref{in:ASI}} 
\and Niccol\`{o} Di Lalla \inst{\ref{in:Stanford}}
\and Immacolata Donnarumma \inst{\ref{in:ASI}}
\and Victor Doroshenko \inst{\ref{in:Tub}}
\and Michal     Dov\v{c}iak \inst{\ref{in:CAS-ASU}}
\and Steven     R. Ehlert \inst{\ref{in:NASA-MSFC}}  
\and Teruaki Enoto \inst{\ref{in:RIKEN}}
\and Yuri~Evangelista \inst{\ref{in:INAF-IAPS}}
\and Riccardo Ferrazzoli \inst{\ref{in:INAF-IAPS}} 
\and Javier     A. Garcia \inst{\ref{in:Caltech}}
\and Shuichi Gunji\inst{\ref{in:Yamagata}} 
\and Kiyoshi Hayashida \inst{\ref{in:Osaka}}\thanks{Deceased}
\and Jeremy Heyl \inst{\ref{in:UBC}}
\and Wataru     Iwakiri \inst{\ref{in:Chiba}} 
\and Svetlana G. Jorstad \inst{\ref{in:BU},\ref{in:SPBU}} 
\and Philip Kaaret \inst{\ref{in:NASA-MSFC}}  
\and Vladimir Karas \inst{\ref{in:CAS-ASU}}
\and Fabian     Kislat \inst{\ref{in:UNH}} 
\and Takao      Kitaguchi  \inst{\ref{in:RIKEN}} 
\and Jeffery J. Kolodziejczak \inst{\ref{in:NASA-MSFC}} 
\and Henric Krawczynski  \inst{\ref{in:WUStL}}
\and Luca Latronico  \inst{\ref{in:INFN-TO}} 
\and Ioannis Liodakis \inst{\ref{in:FINCA}}
\and Simone     Maldera \inst{\ref{in:INFN-TO}}  
\and Alberto~Manfreda \inst{\ref{INFN-NA}}
\and Fr\'{e}d\'{e}ric Marin \inst{\ref{in:Strasbourg}} 
\and Andrea     Marinucci \inst{\ref{in:ASI}} 
\and Alan P. Marscher \inst{\ref{in:BU}} 
\and Herman L. Marshall \inst{\ref{in:MIT}}
\and Francesco~Massaro \inst{\ref{in:INFN-TO},\ref{in:UniTO}} 
\and Ikuyuki Mitsuishi \inst{\ref{in:Nagoya}} 
\and Tsunefumi  Mizuno \inst{\ref{in:Hiroshima}} 
\and Fabio Muleri \inst{\ref{in:INAF-IAPS}} 
\and Michela Negro \inst{\ref{in:UMBC},\ref{in:NASA-GSFC},\ref{in:CRESST}} 
\and Chi-Yung~Ng \inst{\ref{in:HKU}}
\and Stephen L. O'Dell \inst{\ref{in:NASA-MSFC}}  
\and Nicola     Omodei \inst{\ref{in:Stanford}}
\and Chiara     Oppedisano \inst{\ref{in:INFN-TO}}  
\and Alessandro Papitto \inst{\ref{in:INAF-OAR}}
\and George~G.~Pavlov \inst{\ref{in:PSU}}
\and Abel L. Peirson \inst{\ref{in:Stanford}}
\and Matteo     Perri \inst{\ref{in:ASI-SSDC},\ref{in:INAF-OAR}}
\and Melissa Pesce-Rollins \inst{\ref{in:INFN-PI}} 
\and Pierre-Olivier     Petrucci \inst{\ref{in:Grenoble}} 
\and Maura~Pilia \inst{\ref{in:INAF-OAC}} 
\and Andrea Possenti \inst{\ref{in:INAF-OAC}} 
\and Simonetta  Puccetti \inst{\ref{in:ASI-SSDC}}
\and Brian D. Ramsey \inst{\ref{in:NASA-MSFC}}  
\and Ajay Ratheesh \inst{\ref{in:INAF-IAPS}} 
\and Oliver     J. Roberts \inst{\ref{in:USRA-MSFC}}
\and Roger~W.~Romani \inst{\ref{in:Stanford}}
\and Carmelo Sgr\`{o} \inst{\ref{in:INFN-PI}}  
\and Patrick Slane \inst{\ref{in:CfA}}  
\and Paolo Soffitta \inst{\ref{in:INAF-IAPS}} 
\and Gloria     Spandre \inst{\ref{in:INFN-PI}} 
\and Douglas A. Swartz \inst{\ref{in:USRA-MSFC}}
\and Toru~Tamagawa \inst{\ref{in:RIKEN}}
\and Fabrizio Tavecchio \inst{\ref{in:INAF-OAB}}
\and Roberto Taverna \inst{\ref{in:UniPD}} 
\and Yuzuru     Tawara \inst{\ref{in:Nagoya}}
\and Allyn F. Tennant \inst{\ref{in:NASA-MSFC}}  
\and Nicholas~E.~Thomas \inst{\ref{in:NASA-MSFC}}  
\and Francesco  Tombesi  \inst{\ref{in:UniRoma2},\ref{in:INFN-Roma2},\ref{in:UMd}}
\and Alessio Trois \inst{\ref{in:INAF-OAC}}
\and Sergey S. Tsygankov \inst{\ref{in:UTU}}
\and Roberto Turolla \inst{\ref{in:UniPD},\ref{in:MSSL}}
\and Jacco~Vink \inst{\ref{in:Amsterdam}}
\and Martin C. Weisskopf \inst{\ref{in:NASA-MSFC}} 
\and Kinwah     Wu \inst{\ref{in:MSSL}}
\and Fei Xie \inst{\ref{in:GSU},\ref{in:INAF-IAPS}}
          }

   \offprints{M. Cocchi}

\institute{
INAF Osservatorio Astronomico di Cagliari, Via della Scienza 5, 09047 Selargius (CA), Italy  \label{in:INAF-OAC} 
\email{massimo.cocchi@inaf.it}
\and 
Dipartimento di Matematica e Fisica, Universit\`a degli Studi Roma Tre, via della Vasca Navale 84, 00146 Roma, Italy  \label{in:UniRoma3}
\and 
INAF Istituto di Astrofisica e Planetologia Spaziali, Via del Fosso del Cavaliere 100, 00133 Roma, Italy \label{in:INAF-IAPS}
\and 
Department of Physics and Astronomy, FI-20014 University of Turku,  Finland \label{in:UTU} 
\and 
INAF Osservatorio di Astrofisica e Scienza dello Spazio di Bologna, Via P. Gobetti 101, I-40129 Bologna, Italy \label{in:INAF-Bol}
\and 
INAF IASF-Milano, Via Alfonso Corti 12, I-20133 Milano, Italy \label{in:INAF-MI} 
\and 
School of Mathematics, Statistics, and Physics, Newcastle University, Newcastle upon Tyne NE1 7RU, UK \label{in:Newcastle}
\and
Serco for the European Space Agency (ESA), European Space Astronomy Centre, Camino Bajo del Castillo s/n, E-28692 Villanueva de la Ca\~{n}ada, Madrid, Spain \label{in:ESAC}
\and
International Space Science Institute (ISSI), Hallerstrasse 6, 3012 Bern, Switzerland \label{in:ISSI}
\and 
Mullard Space Science Laboratory, University College London, Holmbury St Mary, Dorking, Surrey RH5 6NT, UK \label{in:MSSL}
\and 
Instituto de Astrof\'{i}sicade Andaluc\'{i}a -- CSIC, Glorieta de la Astronom\'{i}a s/n, 18008 Granada, Spain \label{in:CSIC-IAA}
\and 
INAF Osservatorio Astronomico di Roma, Via Frascati 33, 00040 Monte Porzio Catone (RM), Italy \label{in:INAF-OAR} 
\and 
Space Science Data Center, Agenzia Spaziale Italiana, Via del Politecnico snc, 00133 Roma, Italy \label{in:ASI-SSDC}
\and 
Istituto Nazionale di Fisica Nucleare, Sezione di Pisa, Largo B. Pontecorvo 3, 56127 Pisa, Italy \label{in:INFN-PI}
\and  
Dipartimento di Fisica, Universit\`{a} di Pisa, Largo B. Pontecorvo 3, 56127 Pisa, Italy \label{in:UniPI} 
\and 
NASA Marshall Space Flight Center, Huntsville, AL 35812, USA \label{in:NASA-MSFC}
\and  
Istituto Nazionale di Fisica Nucleare, Sezione di Torino, Via Pietro Giuria 1, 10125 Torino, Italy  \label{in:INFN-TO}      
\and  
Dipartimento di Fisica, Universit\`{a} degli Studi di Torino, Via Pietro Giuria 1, 10125 Torino, Italy \label{in:UniTO} 
\and   
INAF Osservatorio Astrofisico di Arcetri, Largo Enrico Fermi 5, 50125 Firenze, Italy 
\label{in:INAF-Arcetri} 
\and  
Dipartimento di Fisica e Astronomia, Universit\`{a} degli Studi di Firenze, Via Sansone 1, 50019 Sesto Fiorentino (FI), Italy \label{in:UniFI} 
\and   
Istituto Nazionale di Fisica Nucleare, Sezione di Firenze, Via Sansone 1, 50019 Sesto Fiorentino (FI), Italy \label{in:INFN-FI}
\and 
Agenzia Spaziale Italiana, Via del Politecnico snc, 00133 Roma, Italy \label{in:ASI}
\and 
Science and Technology Institute, Universities Space Research Association, Huntsville, AL 35805, USA \label{in:USRA-MSFC}
\and 
Istituto Nazionale di Fisica Nucleare, Sezione di Roma ``Tor Vergata'', Via della Ricerca Scientifica 1, 00133 Roma, Italy 
\label{in:INFN-Roma2}
\and 
Department of Physics and Kavli Institute for Particle Astrophysics and Cosmology, Stanford University, Stanford, California 94305, USA  \label{in:Stanford}
\and
Institut f\"ur Astronomie und Astrophysik, Universit\"at T\"ubingen, Sand 1, D-72076 T\"ubingen, Germany \label{in:Tub}
\and 
Astronomical Institute of the Czech Academy of Sciences, Boční II 1401/1, 14100 Praha 4, Czech Republic \label{in:CAS-ASU}
\and 
RIKEN Cluster for Pioneering Research, 2-1 Hirosawa, Wako, Saitama 351-0198, Japan \label{in:RIKEN}
\and 
California Institute of Technology, Pasadena, CA 91125, USA \label{in:Caltech}
\and 
Yamagata University,1-4-12 Kojirakawa-machi, Yamagata-shi 990-8560, Japan \label{in:Yamagata}
\and 
Osaka University, 1-1 Yamadaoka, Suita, Osaka 565-0871, Japan \label{in:Osaka}
\and 
University of British Columbia, Vancouver, BC V6T 1Z4, Canada \label{in:UBC}
\and 
International Center for Hadron Astrophysics, Chiba University, Chiba 263-8522, Japan \label{in:Chiba}
\and
Institute for Astrophysical Research, Boston University, 725 Commonwealth Avenue, Boston, MA 02215, USA \label{in:BU} 
\and 
Department of Astrophysics, St. Petersburg State University, Universitetsky pr. 28, Petrodvoretz, 198504 St. Petersburg, Russia \label{in:SPBU} 
\and 
Department of Physics and Astronomy and Space Science Center, University of New Hampshire, Durham, NH 03824, USA \label{in:UNH} 
\and 
Physics Department and McDonnell Center for the Space Sciences, Washington University in St. Louis, St. Louis, MO 63130, USA \label{in:WUStL}
\and 
Finnish Centre for Astronomy with ESO,  20014 University of Turku, Finland \label{in:FINCA} 
\and 
Istituto Nazionale di Fisica Nucleare, Sezione di Napoli, Strada Comunale Cinthia, 80126 Napoli, Italy \label{INFN-NA}
\and 
Universit\'{e} de Strasbourg, CNRS, Observatoire Astronomique de Strasbourg, UMR 7550, 67000 Strasbourg, France \label{in:Strasbourg}
\and 
MIT Kavli Institute for Astrophysics and Space Research, Massachusetts Institute of Technology, 77 Massachusetts Avenue, Cambridge, MA 02139, USA \label{in:MIT}
\and 
Graduate School of Science, Division of Particle and Astrophysical Science, Nagoya University, Furo-cho, Chikusa-ku, Nagoya, Aichi 464-8602, Japan \label{in:Nagoya}
\and 
Hiroshima Astrophysical Science Center, Hiroshima University, 1-3-1 Kagamiyama, Higashi-Hiroshima, Hiroshima 739-8526, Japan \label{in:Hiroshima}
\and
University of Maryland, Baltimore County, Baltimore, MD 21250, USA \label{in:UMBC}
\and 
NASA Goddard Space Flight Center, Greenbelt, MD 20771, USA  \label{in:NASA-GSFC}
\and 
Center for Research and Exploration in Space Science and Technology, NASA/GSFC, Greenbelt, MD 20771, USA  \label{in:CRESST}
\and 
Department of Physics, University of Hong Kong, Pokfulam, Hong Kong \label{in:HKU}
\and 
Department of Astronomy and Astrophysics, Pennsylvania State University, University Park, PA 16801, USA \label{in:PSU}
\and 
Universit\'{e} Grenoble Alpes, CNRS, IPAG, 38000 Grenoble, France \label{in:Grenoble}
\and 
Center for Astrophysics, Harvard \& Smithsonian, 60 Garden St, Cambridge, MA 02138, USA \label{in:CfA} 
\and 
INAF Osservatorio Astronomico di Brera, via E. Bianchi 46, 23807 Merate (LC), Italy \label{in:INAF-OAB}
\and 
Dipartimento di Fisica e Astronomia, Universit\`{a} degli Studi di Padova, Via Marzolo 8, 35131 Padova, Italy \label{in:UniPD}
\and
Dipartimento di Fisica, Universit\`{a} degli Studi di Roma ``Tor Vergata'', Via della Ricerca Scientifica 1, 00133 Roma, Italy \label{in:UniRoma2}
\and
Department of Astronomy, University of Maryland, College Park, Maryland 20742, USA \label{in:UMd}
\and 
Anton Pannekoek Institute for Astronomy \& GRAPPA, University of Amsterdam, Science Park 904, 1098 XH Amsterdam, The Netherlands  \label{in:Amsterdam}
\and 
Guangxi Key Laboratory for Relativistic Astrophysics, School of Physical Science and Technology, Guangxi University, Nanning 530004, China \label{in:GSU}
}

\date{Received 28 February 2023; accepted May 15 2023}


  \abstract
{After about 16 years since its first outburst, the transient neutron star low-mass X-ray binary \xtej turned on again in September 2022, allowing for the first study of its X-ray polarimetric characteristics by a dedicated observing program with the {\em Imaging X-ray Polarimeter Explorer} (\IXPE).}
{Polarimetric studies of \xtej have been expected to improve our understanding of accreting weakly magnetized neutron stars, 
in particular, the physics and the geometry of the hot inner regions close to the compact object. }
{The \IXPE data of two triggered observations were analyzed using time-resolved spectroscopic and polarimetric techniques, following the source along its Z-track of the color-color diagram.}
{During the first pointing on 2022 September 29, an average 2--8 keV polarization degree of $(4.6\pm 0.4)\%$ was measured, the highest value found up to now for this class of sources. Conversely, only a $\sim$0.6\% average degree was obtained during the second pointing ten days later.}
{The polarimetric signal appears to be strictly related to the higher energy blackbody component associated with the boundary layer (BL) emission and its reflection from the inner accretion disk, and it is as strong as $6.1\%$ and $1.2\%$ ($>95\%$ significant) above 3--4 keV for the two measurements, respectively.
The variable polarimetric signal is apparently related to the spectral characteristics of \xtej, which is the strongest when the source was in the horizontal branch of its Z-track and the weakest in the normal branch. 
These \IXPE results provide new important observational constraints on the physical models and geometry of the Z-sources. Here, we discuss the possible reasons for the presence of  strong and variable polarization among these sources. }


   \maketitle


\section{Introduction}
\label{sec:intro} 

The successful launch in December 2021 of the {\em Imaging X-ray Polarimetry Explorer}~(\IXPE, \citealt{Weisskopf22, Soffitta21}) opened a new investigation territory for the astrophysics of several classes of X-ray astronomical objects. 
In particular, three persistent neutron star low-mass X-ray binaries (NS-LMXBs) were already targeted by \IXPE during its 2022 campaign, namely, the  Z-class source Cyg~X-2 and two bright soft-state Atoll-sources, GS~1826$-$238 and GX~9$+$9. These classes are dubbed after the shape of their X-ray observed hard color versus soft color diagrams \citep[CCDs, see, e.g.,][]{hasinger89,vdK95}.
Cyg~X-2 \citep{Farinelli23} and GX~9+9 \citep{Chatterjee23,Ursini23} have shown a significant linearly polarized component (with the polarization degree, PD, of $\sim2$\%).

\begin{figure}
\centering
\includegraphics[width=0.90\columnwidth]{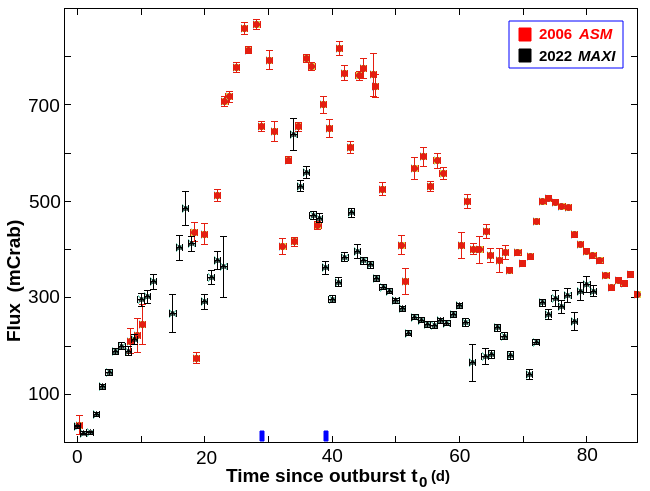}
\caption{2022 (black) and 2006 (red) outburst light curves of XTE J1701-462.
The starting time $t_0$ are MJD 53731.5 and MJD 59824.5 for 2006 and 2022, respectively.
In the instruments' full energy range, 1 Crab is 75 c\,s$^{-1}$ for \XTE/ASM and 3.8 c\,s$^{-1}$ for \MAXI.
The blue ticks on the abscissa indicate the epochs of the two \IXPE observations (no \MAXI data exist during Obs.~1).  
}
\label{fig:lc}
\end{figure}

\xtej can be regarded as a rather unique object and is the first transient NS-LMXB to be observed by \IXPE's modern, sensitive X-ray polarimeters. 
Discovered at the beginning of its 2006 outburst \citep{Remillard06}, the source was soon identified as the seventh Z-source, the first one transient in its nature \citep{Homan06,Homan07}. It also showed kilohertz quasi-periodic oscillations in the \XTE/PCA  data \citep{Homan06b, Homan10, Sanna10} during its Z and atoll (lower banana) phases.
A variable radio counterpart to the X-ray source was identified by \citet{Fender07} along with a very large-scale jet-like multi-blob structure, $\sim$2\arcmin--3\arcmin\ south of the central object, first considered to be connected to a past, unobserved, activity of \xtej, but recently considered unrelated to the source 
\citep{Gasealahwe23}.
The 2006 outburst lasted for a remarkably long time of $\sim 600$\,d \citep{Lin09}, showing an irregular light curve, especially during the long decaying phase \citep{Simon15}. 

The peculiarity of \xtej  is characterized by the fact that is has experienced all the known spectral states of the NS-LMXB during the 2006 outburst \citep{Lin09}.
In the high-soft state (HSS), it was observed in all three Z-branches (horizontal, normal, and flaring branches:\ HB, NB, and FB). Moreover, the source exhibited the typical transient hard tail (THT) usually observed in the HB of the Z-sources \citep{Paizis06}.
Then, \xtej evolved \citep{Homan07b} into a bright soft Atoll source in the Banana spectral state and finally transitioned into a low-hard Atoll Island state.
Up until now, \xtej 
is the only source to have displayed the NS-LMXB complete spectral evolution, also showing the THT spectral component.
This has enabled studies of the physics of different accretion states using timing and spectroscopy in an object-unbiased manner.

Though the NS nature of the compact object in \xtej was already well established, the detection of two type-I X-ray bursts \citep{LvPT93} at the very end of the outburst \citep{Lin09} when the source was in its Island state, was the most solid confirmation.
The bursts had radius expansion characteristics \citep{Lin07} and led to the distance determination of $\sim 8.8$ kpc, using the burst peak luminosity as a standard candle \citep{Kuulkers03}.

A couple of years after its return to quiescence, renewed activity of \xtej was claimed in 2010 by \citet{Krimm10} using the \Swift/BAT data, but not confirmed by \citet{Homan10}.
\xtej turned on again in September 2022 \citep{Iwakiri22}, about 16~yr after its first outburst and about nine months after the beginning of the \IXPE space operations. 
The 2022 outburst looks different in shape and is less luminous than the 2006 one (see Fig.~\ref{fig:lc}). According to the \MAXI light-curve, the source could have returned to quiescence after $\sim 200$ days.
For this second outburst, the X-ray polarimetry 
is
a new important tool to study this NS-LMXB in a unprecedented way.
In the following sections, we present and discuss the results of the \xtej\ \IXPE campaign.


\begin{figure}
\centering
\includegraphics[width=0.8\columnwidth]{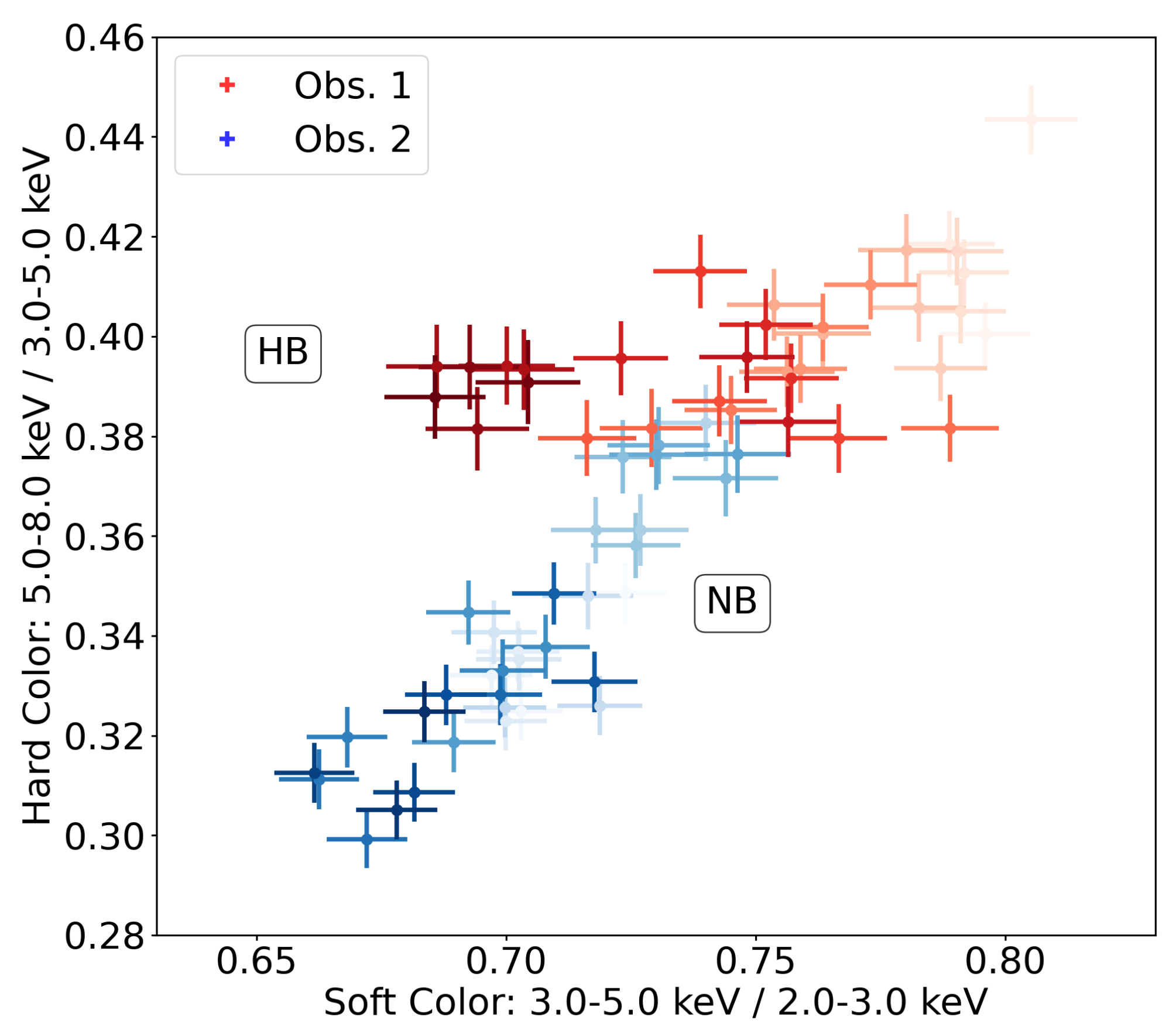}
\caption{\IXPE CCD of \xtej during Obs.~1 (red points) and Obs.~2 (blue points). Colors darken with increasing time in order to show the trend of the patterns. Each point corresponds to 500~s integration time. 
}
\label{fig:ccd}
\end{figure}

\section{\IXPE observations and data analysis}
\label{sec:ixpe} 

The X-ray polarimeter \IXPE \citep{Weisskopf22,Soffitta21} targeted \xtej about 23 days after the \MAXI observed onset of the 2022 outburst on September 6 \citep{Iwakiri22}. 
\IXPE provided high-sensitivity timing, imaging, spectroscopic, and polarimetric of \xtej in the 2--8 keV band.
Two observations (Obs.~1 and 2) were performed on 2022 Sept 29-30 and Oct 8-9 with the nominal exposure of 50~ks each. 
All the three \IXPE detector units (DUs) were fully operational, yielding net exposures of 47.7 ks (DU1, DU2) and 47.2 ks (DU3) for Obs.~1 and 48.1 ks (DU1) and 48.0 ks (DU2,3) for Obs.~2.

Data reduction and analysis were performed using the \textsc{ixpeobssim} software v30.0.1 \citep{Baldini22} and \textsc{heasoft} package version 6.31.1 \citep{2014ascl.soft08004N}. 
The data were filtered by the \textsc{ixpeobssim} tools \texttt{xpselect} and binned with \texttt{xpbin} to produce images and Stokes $I$, $Q,$ and $U$ spectra suitable for spectro-polarimetric analysis with \textsc{xspec}. 
We used the latest (version 12) \IXPE response matrices available on the \textsc{ixpeobssim} public repository.\footnote{https://github.com/lucabaldini/ixpeobssim} 
The source region was selected from the image of each of the three detector units (DU), with the source centred in a circular region of 60\arcsec\ in radius. 
No background subtraction was applied due to the relatively high count rate of the source ($\sim$30 cts~s$^{-1}$ per DU) \citep[see][]{DiMarco23}. 
The data analysis was performed following the weighted method.\footnote{In the weighted analysis method, a weight is assigned to each photo-electron track depending on its shape.} 

The normalized Stokes parameters $q=Q/I$ and $u=U/I$, corresponding to a polarization degree (PD) and angle (PA), as well as their uncertainties were calculated using the model-independent \texttt{pcube} binning algorithm of \textsc{ixpeobssim}.
On the contrary, PD and PA obtained with \textsc{xspec} require the definition of a spectro-polarimetric model. 
The time-resolved spectra were obtained using the procedure described above and then the spectro-polarimetric analysis was performed by \textsc{xspec}.

\begin{figure}
\centering
\includegraphics[width=0.75\columnwidth]{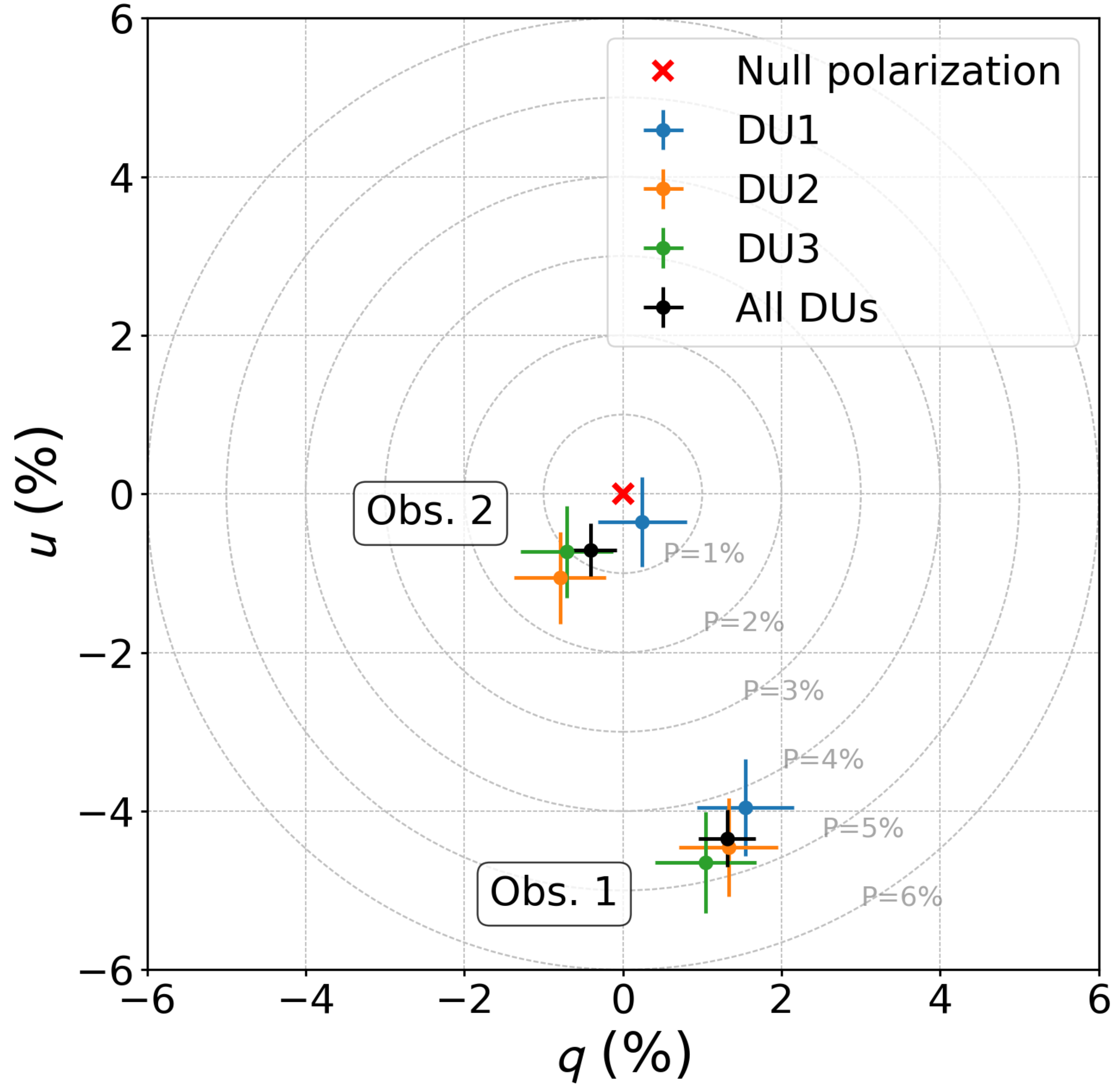}
\caption{Observed 2--8~keV overall polarization signal (black crosses) and for the single \IXPE DUs (colored crosses) for Obs.~1 and Obs.~2. Stokes parameters computed by \texttt{ixpeobsim}.
}
\label{fig:3du}
\end{figure}

\begin{figure}
\centering
\includegraphics[width=0.9\columnwidth]{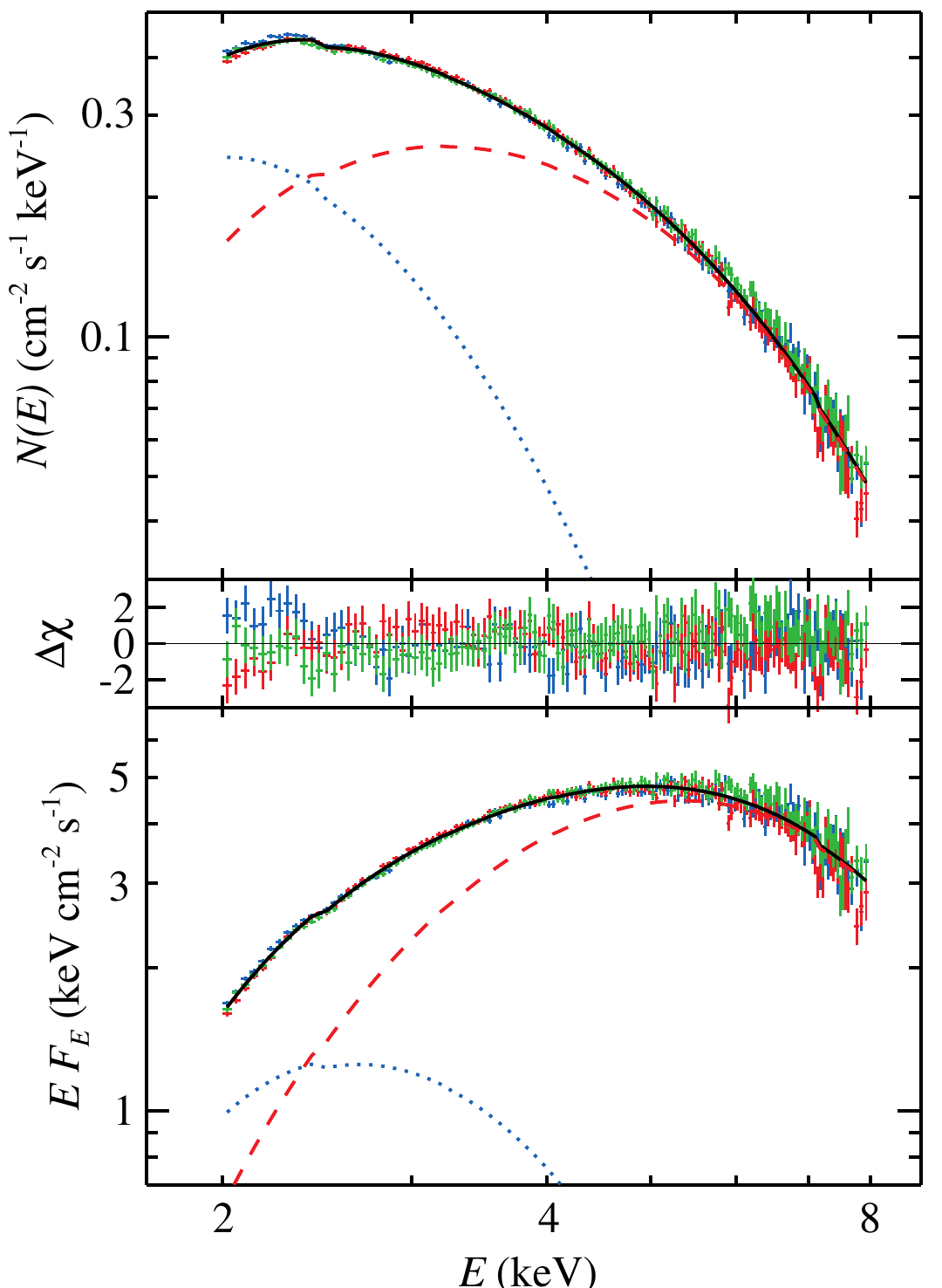}
\caption{\IXPE photon (upper panel) and $EF_E$ spectra (lower panel) of \xtej during Obs.~1, along with the  best-fit model (see Table~\ref{table:1}) consisting of \texttt{diskbb} for the accretion disk (blue dotted curve) and \texttt{bbodyrad} for the BL-SL layer (red dashed) as well as their sum (black solid). The residuals are shown in the middle panel. Blue, red, and green color correspond to DUs 1--3, respectively. 
}
\label{fig:o1_spe}
\end{figure}

\begin{table}
\begin{minipage}[t]{\columnwidth}
\caption{Spectro-polarimetric results for Obs.~1 and 2 for a disk+BL model. 
}
\label{table:1}
\centering
\renewcommand{\footnoterule}{}  

\begin{tabular}{l c c}          
\hline\hline                                    
Parameters  &  Obs.~1 &  Obs.~2\\
\hline
\multicolumn{3}{c}{\texttt{const*tbabs*(diskbb+bbodyrad)} } \\

\texttt{tbabs} & \\
--- $N_{\rm H}$ ($10^{22}~{\rm cm}^{-2}$) &  $2.7\pm0.3$  & $3.4_{-0.3}^{+0.5}$
\\
\texttt{diskbb} & & \\ 
--- $T_{\rm in}$~~(keV)   &    $0.72_{-0.07}^{+0.09}$  & $0.57\pm0.04$ \\
--- $R_{\rm in}\ \sqrt{\cos i}$ (km)\tablefootmark{a}  &  $37_{-12}^{+16}$ & $78_{-23}^{+35}$
\\
\texttt{bbodyrad}  & & \\
--- $T$~~(keV)   &  $1.31_{-0.02}^{+0.03}$ & $1.13\pm0.01$  \\
--- $R_{\rm bb}$  (km)   &  $18\pm3$ & $26_{-3}^{+4}$
\\
\texttt{const} & & \\
--- DU1      &  $[ 1 ]$  & $[ 1 ]$ \\
--- DU2      &  $0.981 \pm 0.003$ & $0.967 \pm 0.003$ \\
--- DU3      &  $0.917 \pm 0.003$ & $0.950 \pm 0.004$ \\
\\
Reduced $\chi^{2}$ (d.o.f)    &  1.044 (440)  &  1.080 (422) \\
\\
Flux ($10^{-9}$\,erg\,cm$^{-2}$\,s$^{-1}$)  & $8.34_{-0.62}^{+0.01}$  & $8.73_{-0.54}^{+0.02}$ \\  
Photon flux (cm$^{-2}$\,s$^{-1}$)  & $1.304_{-0.125}^{+0.001}$  & $1.417_{-0.114}^{+0.004}$ \\  
disk photons\tablefootmark{b} & 0.247   & 0.226 \\
\hline
\\
\multicolumn{3}{c}{\texttt{polconst*(diskbb+bbodyrad)}\tablefootmark{c}  } \\

--- PD (\%)     & $4.58 \pm 0.37$& $0.65 \pm 0.35$ \\
--- PA  (deg)   & $-37.7 \pm 2.3$ & $-57 \pm 16$ 
\\ \\
\multicolumn{3}{c}{\texttt{(polconst*diskbb + polconst*bbodyrad)}\tablefootmark{d} } \\

\texttt{diskbb} & & \\ 
--- PD (\%)   & $< 1.6$ & $< 3.2$ \\
--- PA (deg) & [PA$_{\rm BL}$+90] & [PA$_{\rm BL}$+90] \\
\texttt{bbodyrad} (for BL) & & \\ 
--- PD (\%)   & $6.14_{-0.49}^{+0.69}$ & $1.22 \pm 0.65$\\
--- PA (deg)   & $-38.3 \pm 2.3$ & $-54 \pm 14$ \\
\texttt{bbodyrad}\tablefootmark{e} & & \\ 
--- PD (\%)   & $6.16_{-0.50}^{+0.52}$ & $0.90\pm 0.45$ \\
--- PA (deg)  & $-38.3 \pm 2.3$ & $-56 \pm 15$ \\
\hline
\end{tabular}
\tablefoot{DU energy range is 2--8 keV. 
The spectral and polarimetric fits errors and upper limits are joint computed at $2 \sigma$ (95\% confidence).
A systematics of 1\% is applied in \xspec in order to account for slight instrumental miscalibration.
Normalizations of DU2 and DU3 are relative to DU1. 
Frozen or linked parameters are in square brackets. 
\tablefoottext{a}{Radia are determined for the distance of 10 kpc.}
\tablefoottext{b}{Fraction of the total photon flux.}
\tablefoottext{c}{Spectroscopic parameters were first set to their best-fit values (upper panel) and then allowed to vary in the spectro-polarimetric fit.} 
\tablefoottext{d}{The disk PA is set $90\degr$ apart from the BL one.}
\tablefoottext{e}{As above, but freezing [PD=0] for the disk component.}
}
\end{minipage}
\end{table}

\section{Results}
\label{sec:results}

We constructed the time-resolved CCD for both observations (see Fig.~\ref{fig:ccd}).
During Obs.~1, the source moved from a lower HB to an upper HB diagram position.
This coincides with the 2-8 keV time profile flux decreasing by $\sim 28\%$, in an anti-correlation to its 4--8/2--4 keV X-ray hardness ratio, as is typical for a Z-source in the HB. 
Furthermore,Obs.~2 found the source in its NB, moving from the upper NB to the lower NB.
During the observation, the flux and the hardness ratio were stable within a few percent, in agreement with a NB where the source is expected to be very close to its Eddington-limited luminosity.

\begin{figure} 
\centering
\includegraphics[width=0.85\columnwidth]{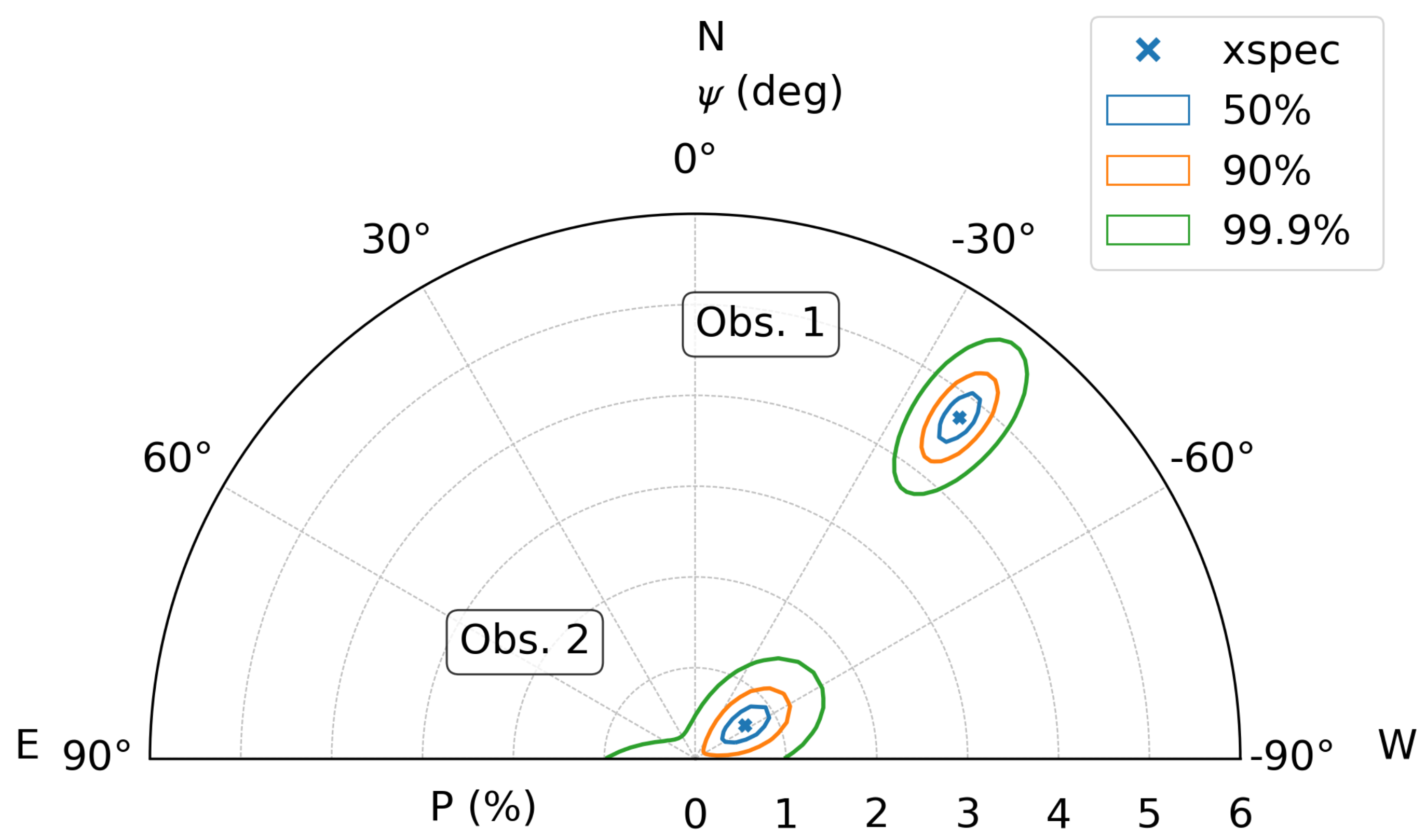}
\caption{Average polarization 
for Obs.~1 and 2 computed via spectro-polarimetric analysis using a single \texttt{polconst} convolving component in \xspec (see Table~\ref{table:1}). 
}
\label{fig:xO1O2}
\end{figure}

The \IXPE polarimetric results obtained by \texttt{ixpeobsim} showed a (model-independent) strong variability of the linear PD between the two observations (see Fig.~\ref{fig:3du}).
An average PD of $(4.6\pm 0.4)\%$ was found during Obs.~1, while in Obs.~2 ($\sim 10$~d after Obs.~1), the 2--8 keV polarimetric signal could not be constrained with the 99\% upper limit of 1.5\%.

We described the time averaged spectra of the two 
observations by a standard two-component HSS model consisting of a multicolor disk and a harder boundary-spreading layer (BL-SL) emission \citep[e.g.,][]{Popham01, Revnivtsev13}.
The BL emission can be modeled in \xspec either using a pure blackbody (\texttt{bbodyrad}), or an unsaturated Comptonization model, such as  \texttt{comptt} \citep{Titarchuk94}.
The wide-band measurements (e.g. with \SAX) suggest a Comptonized emission from a BL \citep[e.g.][]{disalvo02,Farinelli09} as a better way to describe the data. 
However, the \IXPE spectra of \xtej turned out difficult to fit by a \texttt{comptt} component without additional assumptions, due to a degeneracy in the parameter space.
This is most likely due to the relatively narrow data pass-band, which prevents the energy cutoff from being constrained.
Therefore, we considered replacing in the fits the Comptonization model by a pure blackbody component. In this case, the fits converged without further constraints.

The disk emission was modeled by \texttt{diskbb} and \texttt{diskpn} \citep{mitsuda84,gierlinski99}.
Due to the increased parameter space for \texttt{diskpn}, the inner radius resulted unconstrained, so it could not be distinguished from \texttt{diskbb}, and, in fact, the temperatures of the two models are consistent with each other.

A \texttt{tbabs} \citep{Wilms00} multiplicative component was included to account for the low-energy absorption due to the interstellar medium;
abundances and cross-section tables according to \cite{Asplund09} and \cite{Verner96}, (\texttt{aspl, vern} in \xspec).
We also included a normalizing cross-calibration multiplicative factor to account for the uncertainties in the effective areas of three telescopes.
The resulting best fits for both observations are reported in Table~\ref{table:1} (upper panel) for the disk plus the blackbody-modeled BL.
Photon and energy fluxes are slightly higher in Obs.~2, in agreement with an expected larger energy production in the NB.
The BL and the inner disk are found to be hotter in Obs.~1, while the inner disk radius seems larger in Obs.~2. The opposite occurs for the BL sphere-equivalent radius.
We tested the reasonable hypothesis that $N_{\rm H}$  and the normalizing factors were the same for both observations. Nevertheless, the corresponding fits could be accepted only by relaxing the systematics to 1.3\% and 1.5\%. 
On the other hand, the interpolated value (\texttt{nh} ftool, v.3) of $N_{\rm H}$ at the source coordinates is $\sim 0.6\times10^{22} {\rm cm}^{-2}$, namely, it is several times lower than those presented in \citet{Lin09} and in the present work. This suggests the source may be intrinsically (and possibly variably) absorbed by nearby cold structures.

The Obs.~1 spectral best fit is shown in Fig.~\ref{fig:o1_spe}. 
The softer disk component dominates in photons (upper panel) up to $\sim 2.5$~keV, being still important up to $\sim 3.5$~keV.
In the 2--8~keV band, the disk photons account for $\sim 24\%$ of the total. 
The harder BL completely dominates the energy budget (lower panel, $EF_E$ representation).

\begin{figure}
\centering
\includegraphics[width=0.9\columnwidth]{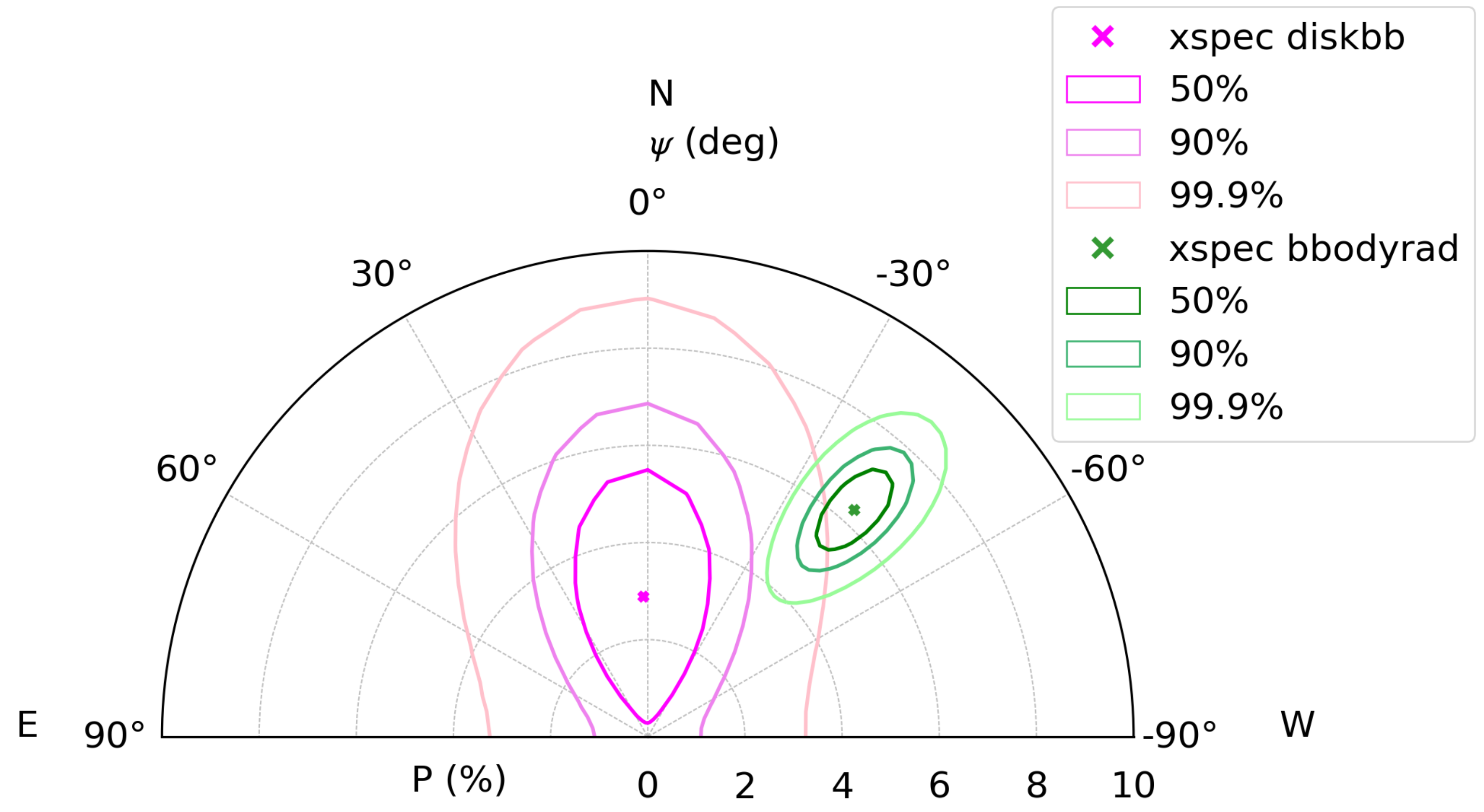}
\caption{Polarization contours in Obs. 1 for the \texttt{diskbb} and \texttt{bbodyrad} components estimated with \texttt{polconst} (see also Table 1, lower fit).
}
\label{fig:obs1comp2}
\end{figure}

Concerning the average polarization in the two observations,
fully \texttt{ixpeobsim}-compatible results were achieved by the spectral-polarimetric data fits using \xspec 
(please compare Fig.~\ref{fig:xO1O2} to Fig.~\ref{fig:3du}), where the energy-independent polarized multiplicative model component (\texttt{polconst}) was used (see Table~\ref{table:1}).

Motivated by the spectral decomposition shown in Fig.~\ref{fig:o1_spe}, we performed a spectro-polarimetric analysis, assuming that the two components may have different polarization properties.
In Table~\ref{table:1}, lower panel, the average and disk-versus-BL resolved (the latter obtained by applying a dedicated \texttt{polconst} to each of the main spectral components) polarization results are reported. 
The BL-related signals were constrained in both observations, with a polarization degree of 6.1\% and 1.2\%, respectively. In Obs.~2 the signal is incompatible with null polarization with a $>$99.7\% probability.
The best fits and errors were obtained by joint variation of the spectral and polarimetric parameters \citep{Rawat23} starting from the best fit to the spectra shown in the upper part of  Table~\ref{table:1}.
As in \citet{Farinelli23}, we assumed a 90\degr\ separation of the disk polarization angle (PA) with respect to the BL one. This is expected in the absence of strong GR effects and because of the different geometries of the scattered BL radiation and the intrinsically polarized disk radiation (e.g., \citet{Chandrasekhar60}).
We also investigated the possibility of fully independent PAs for the two components, but no conclusion could be drawn because of the unconstrained and not statistically significant
disk signal (see also Fig.~\ref{fig:obs1comp2} for Obs.~1).

\begin{figure}
\centering
\includegraphics[width=0.95\columnwidth]{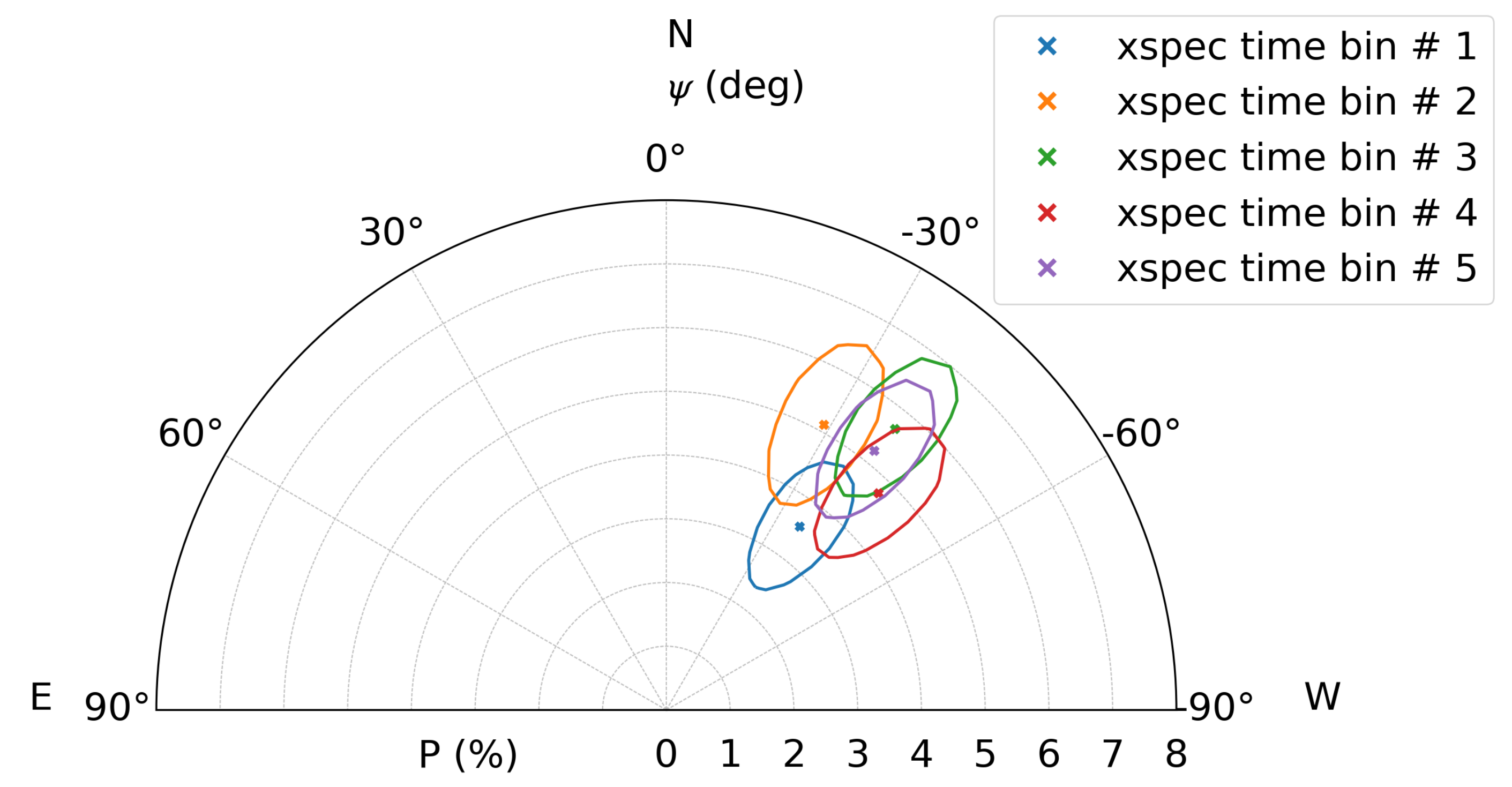} 
\caption{Time-binned polarization of Obs.~1 at 90\% confidence level. Zero polarization is excluded for all time bins at 99.9\% confidence level. 
}
\label{fig:obs1obs2timebins}
\end{figure}

\begin{figure}
\centering
\includegraphics[width=0.38\columnwidth]{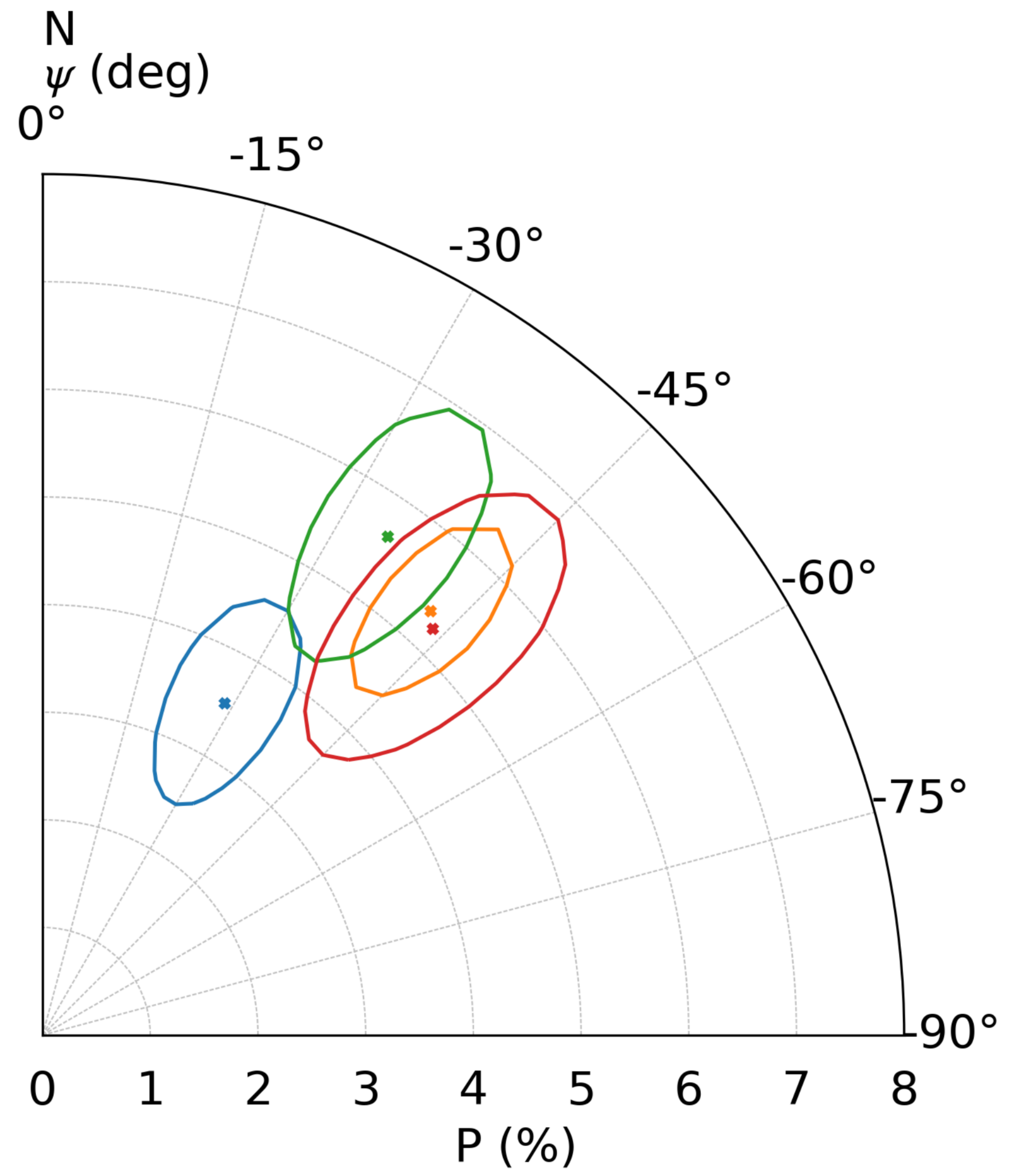}
\includegraphics[width=0.61\columnwidth, trim = 0 0 -200 0]{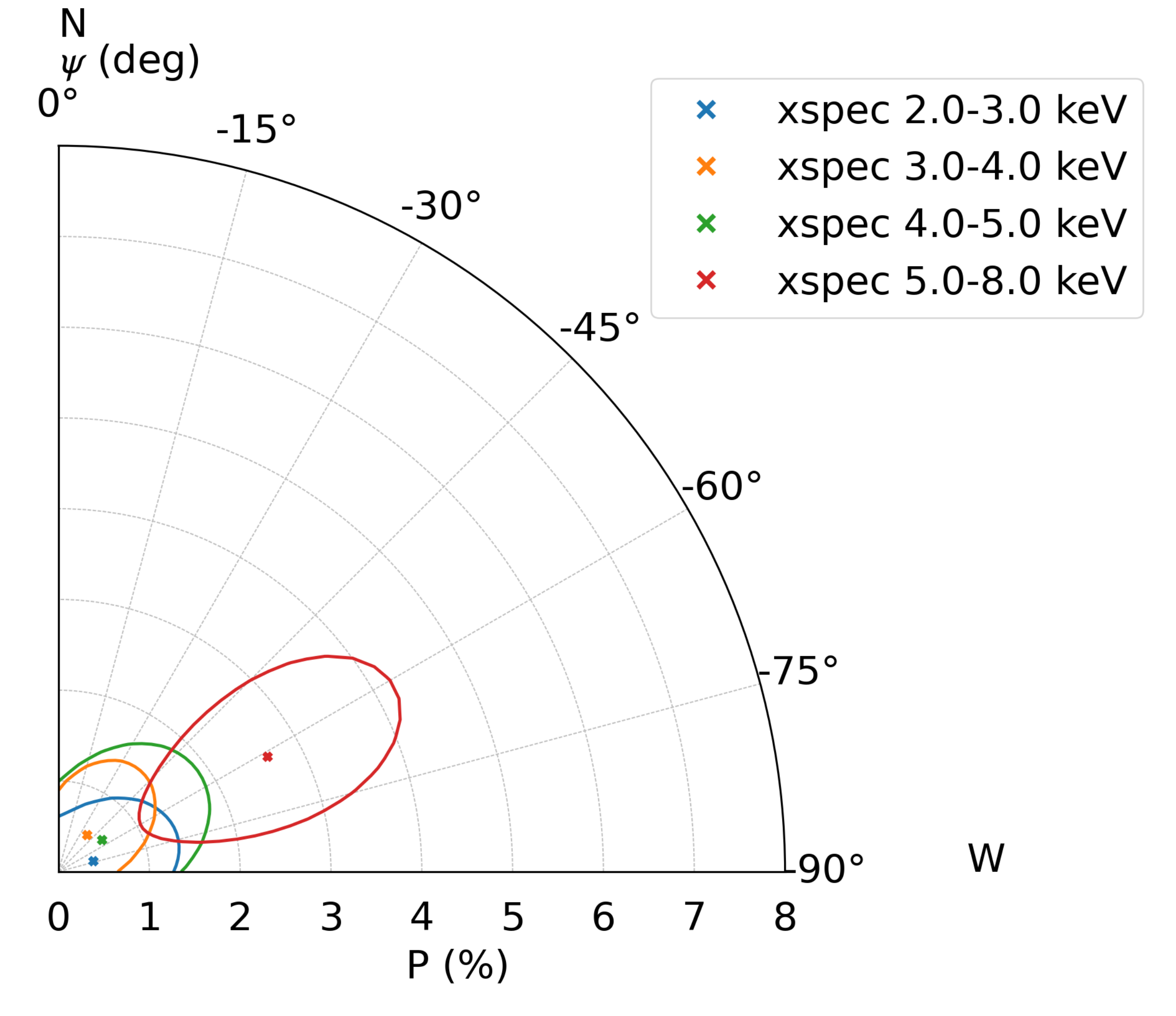}
\caption{90\% confidence level contours for the PD and PA in four \IXPE energy bands  (2--3, 3--4, 4--5, and 5--8 keV) for Obs.~1 (left panel) and Obs. 2 (right panel).
}
\label{fig:etrend}
\end{figure}

The strong average polarization signal during Obs.~1 does not show a clear evidence for time variability during the measurement, although
the polarization time profile (obtained in five 10 ks time intervals)\ may suggest a slightly lower average signal in the first time bin
(see Fig.~\ref{fig:obs1obs2timebins}).
The energy dependence of the polarization during both observations is not evident either (as shown in Fig.~\ref{fig:etrend}).
Nevertheless, the lowest energy (2--3~keV) bin in Obs.~1 shows a lower polarization at a 90\% confidence level with respect to higher energy bins. 

In Obs.~2, only the 5--8~keV energy bin is compatible with a non-null polarization, at a confidence of 99.73\% .
If we assume Obs.~2 is unpolarized in all the four independent energy bins, the probability of claiming a false polarization for the 5--8~keV bin is $1-(1-p)^n$, where $p=0.27\%$ (for the single-bin confidence level of  99.73\%) and $n=4$.
This probability is $1.08\%$. 
The 5-8~keV polarization, being roughly aligned with the signal of Obs.~1, supports the possibility of a (weak) BL-related signal in Obs.~2 as well.

\section{Discussion}
\label{sec:disc}

The results in Section~\ref{sec:results} show that \xtej was observed by \IXPE during its outburst Z evolutionary phase, in agreement with the 
reports of other X-ray observatories \citep[e.g.,][]{Thomas22}. 
The \IXPE light curve, CCD diagrams, and spectroscopic analysis confirm this result with a high level of confidence.
In this context, the discovery of a strong polarimetric signal during the 2022 Sept 29 observation (Obs.~1) poses a new important constraint on the physical models and the geometry of the emission region of Z-sources.

The relative fluxes of the disk and the BL during Obs.~1 (Fig.~\ref{fig:o1_spe}) are in agreement with those typically observed for Z-sources in their HB CCD track \citep{Farinelli09}. Also, the CCD time evolution  (Fig.~\ref{fig:ccd}) during the observation indicates that the source was in a HB of the Z-track.
Conversely, according to the CCD and the average spectral characteristics, Obs.~2 most likely found the source in the NB of the Z-track, with no (or relatively insignificant) excursions to the HB; in this case, 
only a $\sim 1\%$ polarized signal that is strictly connected to the BL component could be established.
This indicates that polarization is connected to the source position along its Z-track, and it is strongest in the upper-intermediate HB.

Taking these findings into account, it is necessary to explain the evidence for strong (possibly transient) polarization in the upper HB of the Z-track.
Actually, the measured PD of $\sim 6\%$ associated with the BL (direct and/or disk-reflected) emission as observed from a Z-source in a HB is rather challenging to explain. 

An electron-scattering dominated accretion disk can be polarized at a level of 4\% \citep{Chandrasekhar60} for the inclination of $i=75\degr$ as inferred for \xtej by \citet{Lin09}.
However, the disk dominates below $\sim 3$\,keV and so, it cannot be responsible for the polarized emission at higher energies where the BL dominates.
The BL itself, as a geometrically thicker extension of the inner accretion disk (i.e. lying in the same plane) is not likely to be the source of the polarized signal.
Achieving PD of 5\% requires a high inclination $>78\degr$ \citep{Chandrasekhar60}, typical for a dipper, while \xtej does not show dips. Moreover, in a very similar source, Cyg X-2, polarization is parallel to the jet direction \citep{Farinelli23}, while the BL model would predict polarization perpendicular to the jet. 
A high polarization of $\sim 10\%$ can be easily achieved by multiple Compton scatterings in an optically thin, plane-parallel hot corona above the disk \citep{PS96}, however, this spectral component is not observed in the \IXPE data. 

The spreading layer (SL) \citep{inogamov1999}, namely, the BL at the NS surface, can potentially be polarized \citep{Lapidus85}. 
We have computed the expected PD for this model assuming that the local emission corresponds to the semi-infinite electron-scattering dominated atmosphere \citep{Chandrasekhar60}.
We considered layers of different latitudinal extents, narrow belts, and fully covered NS, and we accounted for the relativistic effects related to the rotation of the polarization plane due to the fast motion of the SL \citep{poutanen2020}. 
We did not obtain a PD exceeding $\sim 1.5\%   $ in any of these cases. 
Thus, we can conclude that the SL alone is not responsible for the observed polarization. 

However, radiation that is emitted by the SL at the NS surface may be scattered by the surrounding material. 
If the SL does not extend to very high latitudes, then reflection from the ionized inner accretion disk could produce PD up to 6\% in the total spectrum \citep[see Fig. 7 in][]{Lapidus85} parallel to the rotation axis, when neglecting the inner disk direct emission (which is most likely polarized, but counteracting as being perpendicular to the axis).
However, it is also important to note that disk polarization drops very rapidly with decreasing inclination, while the PD of the reflected signal has a much slower dependence on inclination. 
This implies that the PD of the total signal will be somewhat smaller at lower energies due to the dilution by nearly unpolarized disk radiation, but will not drop to zero due to cancellation of perpendicular polarization.

Radiation from the SL can also be scattered in a wind from the disk -- or a jet, if one is present in the system. 
Radio loudness is mostly reported when the Z-sources are found in their HB phase \citep[e.g.,][]{DGK07}. 
Moreover, the observations of transient hard tails (THT) in the spectra of Z-sources are strictly correlated to HB positions in their CCDs \citep{damico2001, disalvo02, iaria2004,farinelli2005} and possibly produced by scattering in a jet-like outflow \citep{Reig16}.

It is thus tempting to link the HB strong polarimetric signal to the accelerating mechanism (still not fully understood) that produces the jet emission and the possibly jet-related THT \citep{Reig16}. 
However, the contribution of the THT to the flux in the \IXPE band is not expected to exceed 10\% \citep[see, e.g.,][]{Paizis06} and therefore it is not very likely that this component is responsible for a strong polarized signal, unless the BL photons are single-scattered in the Thomson regime.
Furthermore, it is not clear why the PD would then drop at lower energies, unless the spectrum of the THT has a low-energy cutoff around 3-4 keV; however, this can be nevertheless expected because a power-law component with no low energy cutoff would cause a low-energy fit contamination, which is not observed in the wide band spectra of Z-sources during their THT spectral phase \citep{Paizis06,Farinelli09}.

As the previously claimed extended jet-like structure to the south of \xtej \citep{Fender07} is more likely a background feature \citep{Gasealahwe23}, there are no convincing arguments as to why the jet would be the source of polarized signal.
The central source itself, however, was radio loud \citep{Fender07}, implying the existence of some outflow from the system. Also, the similarity of \xtej to the prototype Cyg~X-2 suggests that polarimetric signal is likely parallel to the outflow.
The scattering of central radiation in the outflow can produce polarization.  
If the outflow occupied a large area above the disk, both the emission of the SL and the disk should be polarized in a similar fashion, which is not the case.

On the other hand, the outflow originating from the inner part of the accretion disk may face much larger optical depth along the disk plane than perpendicular to it. 
Thus, the disk radiation which is beamed strongly along its normal will be affected very little, whereas the SL radiation, which is emitted more isotropically (or even slightly beamed along the disk surface) can be scattered much more efficiently. 
Radiation injected along a slab and Thomson scattered there gets polarized parallel to the slab normal with the PD$\,=\sin^2i/(3-\cos^2i)$ \citep{st85}, reaching the maximum of
33\% at 90\degr\ inclination. 
The PD in this case has a rather weak dependence on the inclination and even at $i=70\degr$, the PD is still $\sim 30\%$.
The observed 5--6\% polarization requires that about 20\% of the SL radiation is scattered in the wind, which does not appear to be implausible.

Finally, the question of why the PD is much smaller in Obs.~2 must be addressed.
If scattering in the wind is responsible for the high PD in Obs.~1, then the wind has to almost disappear -- or, at least, its optical depth should become smaller in the NB. 
If reflection from the disk is the source of the high PD, the most straightforward explanation would be that the latitudinal extent of the SL has decreased during Obs.~2, so that it no longer illuminates the disk efficiently. 

Our \IXPE data (see Table~\ref{table:1}) also suggest that the inner disk could be farther from the SL during Obs.~2, possibly due to very high radiation pressure (Z-sources reach their Eddington limit when in the NB). This fact also points to less efficient disk illumination.
Of course, all these speculations need the support of reliable simulations and this will likely be the subject of future investigations.

\section{Summary}

We reported a $\sim 5\%$ linearly polarized signal in \IXPE observations of the NS transient Z-source \xtej. 
The signal is found when the source is at the upper-and-intermediate horizontal branch position in the CCD diagram.  
The unprecedentedly high strength of the signal for this class of compact objects can hardly be explained by a standard boundary layer plus accretion disk scenario.  
We speculate that scattering of the boundary-spreading layer radiation by either the accretion disk, a corona above the disk, or an optically thin wind can be responsible for the observed polarimetric characteristics. 
On the other hand, the polarimetric signal was much lower when the source was in the normal branch, implying a change in the accretion geometry or the wind properties.

\begin{acknowledgements}
The Imaging X-ray Polarimetry Explorer (IXPE) is a joint US and Italian mission.  The US contribution is supported by the National Aeronautics and Space Administration (NASA) and led and managed by its Marshall Space Flight Center (MSFC), with industry partner Ball Aerospace (contract NNM15AA18C).  
The Italian contribution is supported by the Italian Space Agency (Agenzia Spaziale Italiana, ASI) through contract ASI-OHBI-2017-12-I.0, agreements ASI-INAF-2017-12-H0 and ASI-INFN-2017.13-H0, and its Space Science Data Center (SSDC) with agreements ASI-INAF-2022-14-HH.0 and ASI-INFN 2021-43-HH.0, and by the Istituto Nazionale di Astrofisica (INAF) and the Istituto Nazionale di Fisica Nucleare (INFN) in Italy.
This research used data products provided by the IXPE Team (MSFC, SSDC, INAF, and INFN) and distributed with additional software tools by the High-Energy Astrophysics Science Archive Research Center (HEASARC), at NASA Goddard Space Flight Center (GSFC). 
J.P. and A.V. acknowledge support from the Academy of Finland grant 333112. A.B. is supported by the Finnish Cultural Foundation grant 00220175.
The authors also thank the anonymous referee for the useful comments and suggestions.
\end{acknowledgements}


\end{document}